\documentclass[letterpaper,11pt]{article}

\usepackage[width=172mm,height=224mm]{geometry}
\usepackage{amsmath}
\usepackage{amsthm}
\usepackage{amssymb}
\usepackage{mathtools}
\usepackage{times}
\usepackage[T1]{fontenc}
\usepackage{url}
\usepackage{enumerate}

\newcommand{\NP}{\mathrm{NP}}
\newcommand{\PSPACE}{\mathrm{PSPACE}}
\newcommand{\EXP}{\mathrm{EXP}}
\newcommand{\NEXP}{\mathrm{NEXP}}
\newcommand{\IP}{\mathrm{IP}}
\newcommand{\MIP}{\mathrm{MIP}}
\newcommand{\MIPns}{\mathrm{MIP}^{\mathrm{ns}}}
\newcommand{\QMIPns}{\mathrm{QMIP}^{\mathrm{ns}}}
\newcommand{\QIP}{\mathrm{QIP}}
\newcommand{\QRG}{\mathrm{QRG}}
\newcommand{\poly}{\mathrm{poly}}

\DeclarePairedDelimiter{\abs}{\lvert}{\rvert}

\newcommand{\RR}{\mathbb{R}}
\newcommand{\ZZ}{\mathbb{Z}}

\newcommand{\wns}{w_{\mathrm{ns}}}

\theoremstyle{plain}
\newtheorem{theorem}{Theorem}
\newtheorem{lemma}[theorem]{Lemma}
\newtheorem{corollary}[theorem]{Corollary}
\newtheorem{claim}{Claim}
\theoremstyle{remark}
\newtheorem{remark}{Remark}

\newcommand{\Ref}{}
\newcommand{\Refs}{}

\title{Polynomial-Space Approximation of No-Signaling Provers}
\author{Tsuyoshi Ito \\
  \normalsize tsuyoshi@iqc.ca \\
  \normalsize Institute for Quantum Computing and School of Computer Science \\
  \normalsize University of Waterloo}
\date{October~20, 2009}

\begin{document}
\maketitle

\begin{abstract}
  In two-prover one-round interactive proof systems,
  no-signaling provers are those who are allowed to use arbitrary strategies,
  not limited to local operations,
  as long as their strategies cannot be used for communication between them.
  Study of multi-prover interactive proof systems with no-signaling provers
  is motivated by study of those with provers sharing quantum states.
  The relation between them
  is that no-signaling strategies include all the strategies
  realizable by provers sharing arbitrary entangled quantum states, and more.

  This paper shows that two-prover one-round interactive proof systems
  with no-signaling provers
  only accept languages in $\PSPACE$.
  Combined with the protocol for $\PSPACE$
  by Ito, Kobayashi and Matsumoto (CCC 2009),
  this implies $\MIPns(2,1)=\PSPACE$,
  where $\MIPns(2,1)$ is the class of languages
  having a two-prover one-round interactive proof system
  with no-signaling provers.
  This is proved by constructing a fast parallel algorithm
  which approximates within an additive error
  the maximum value of a two-player one-round game
  achievable by cooperative no-signaling players.
  The algorithm uses the fast parallel algorithm
  for the mixed packing and covering problem by Young (FOCS 2001).
\end{abstract}

\section{Introduction}

\subsection{Background}

Nonlocality~\cite{Bell64Physics} is a peculiar property of quantum mechanics
and has applications to quantum information processing.
Following Cleve, H\o yer, Toner and Watrous~\cite{CleHoyTonWat04CCC},
quantum nonlocality can be naturally expressed
in terms of cooperative two-player one-round game with imperfect information,
which is a game played by two players and a referee as follows.
The players are kept in separate rooms so that they cannot communicate with each other.
The referee chooses a pair of questions according to some probability distribution,
and sends one question to each player.
Each player replies with an answer to the referee,
and the referee declares whether the two players jointly win or jointly lose
according to the questions and the answers.
The players know the protocol used by the referee
including the probability distribution of the pair of questions
and how the referee determines the final outcome of the game,
but none of the players knows the question sent to the other player.
The aim of the players is to win the game with as high probability as possible,
and the maximum winning probability is called the \emph{value} of the game.
In this framework, a Bell inequality is an inequality
stating an upper bound of the value of a game of this kind
when provers are not allowed to perform any quantum operations,
and the violation of a Bell inequality means that
the game value increases
when provers are allowed to share a quantum state before the game starts.

The complexity of finding or approximating the value of a game
has been one of the most fundamental problems
in computational complexity theory.
The computational model based on cooperative multi-player games
is called multi-prover interactive proof systems
and were introduced
by Ben-Or, Goldwasser, Kilian and Wigderson~\cite{BenGolKilWig88STOC}
for a cryptographic purpose.%
\footnote{Because of this connection,
  we use ``player'' and ``prover'' synonymously in this paper.}
It turned out that this computational model is extremely powerful:
multi-prover interactive proof systems
exactly characterize $\NEXP$~\cite{ForRomSip94TCS,BabForLun91CC},
even in the most restricted settings
with two provers, one round and an exponentially small one-sided error~%
\cite{FeiLov92STOC}.
In other words, given the description of a cooperative game,
approximating the best strategy even in a very weak sense
is notoriously difficult.
These results were built on top of techniques developed
in studies on (single-prover) interactive proof systems~%
\cite{Babai85STOC,GolMicRac89SICOMP,LunForKarNis92JACM,Shamir92JACM}
as well as multi-prover interactive proof systems with weaker properties~%
\cite{CaiConLip94JCSS,Feige91SCT,LapSha97JCSS}.
It is noteworthy that the powerfulness of multi-prover one-round interactive proof systems
has led to a successful study of probabilistically checkable proof systems~%
\cite{BabForLevSze91STOC,FeiGolLovSafSze96JACM},
which play a central role
in proving $\NP$-hardness of many approximation problems
via the celebrated PCP theorem~\cite{AroSaf98JACM,AroLunMotSudSze98JACM}.

Cleve, H\o yer, Toner and Watrous~\cite{CleHoyTonWat04CCC}
connected the computational complexity theory and the quantum nonlocality
and raised the question on the complexity of approximating
the value of a cooperative game with imperfect information
in the case where the players are allowed to share quantum states
or, in terms of interactive proof systems,
the computational power of multi-prover interactive proof systems
with entangled provers.
Kobayashi and Matsumoto~\cite{KobMat03JCSS}
considered another quantum variation of multi-prover interactive proof systems
where the verifier can also use quantum information
and can exchange quantum messages with provers,
which is a multi-prover analogue of quantum interactive proof systems~\cite{Watrous03TCS}.
In \Ref\cite{KobMat03JCSS},
it was shown that allowing the provers to share at most polynomially many qubits
does not increase the power of multi-prover interactive proof systems
beyond $\NEXP$
(even if the verifier is quantum).
Although studied intensively~%
\cite{KemKobMatTonVid08FOCS,CleGavJai09QIC,ItoKobPreSunYao08CCC,KemRegTon08FOCS,KemKobMatVid09CC,
      Gutoski09QIC,DohLiaTonWeh08CCC,NavPirAci08NJP,BenHasPil08FOCS,ItoKobMat09CCC},
the power of multi-prover interactive proof systems
with provers allowed to share arbitrary quantum states
has been still largely unknown.

The notion of no-signaling strategies
was first studied in physics in the context of Bell inequalities
by Khalfin and Tsirelson~\cite{KhaTsi85SFMP} and Rastall~\cite{Rastall85FP},
and it has gained much attention
after reintroduced by Popescu and Rohrlich~\cite{PopRoh94FP}.
The acceptance probability of the optimal no-signaling provers
is often useful as an upper bound of the acceptance probability of entangled provers
(and even commuting-operator provers
 based on the notion of commuting-operator behaviors;
 see~\cite{Tsirelson06,NavPirAci08NJP,DohLiaTonWeh08CCC,ItoKobPreSunYao08CCC})
because no-signaling strategies have a simple mathematical characterization.
Toner~\cite{Toner09PRSA} uses no-signaling provers
to give the maximum acceptance probability of entangled provers
in a certain game.
Extreme points of the set of no-signaling strategies
are also studied~\cite{BarLinMasPirPopRob05PRA,AviImaIto06JPhysA}.

Kempe, Kobayashi, Matsumoto, Toner and Vidick~\cite{KemKobMatTonVid08FOCS} prove,
among other results, that every language in $\PSPACE$
has a two-prover one-round interactive proof system
which has one-sided error $1-1/\poly$
even if honest provers are unentangled and dishonest provers are allowed
to have prior entanglement of any size
(the proof is in \Ref\cite{KemKobMatTonVid-0704.2903v2}).
Ito, Kobayashi and Matsumoto~\cite{ItoKobMat09CCC} improve their result
to an exponentially small one-sided error by considering no-signaling provers;
more specifically, they prove that the soundness of the protocol
in \Ref\cite{KemKobMatTonVid08FOCS}
actually holds against arbitrary no-signaling provers,
then use the parallel repetition theorem
for no-signaling provers~\cite{Holenstein09TOC}.
We note that the soundness analysis of \Ref\cite{ItoKobMat09CCC}
is somewhat simpler than that of \Ref\cite{KemKobMatTonVid08FOCS}.

Repeating the protocol of \Ref\cite{KemKobMatTonVid08FOCS} parallelly
as is done in \Ref\cite{ItoKobMat09CCC}
results in the protocol identical to the one
used by Cai, Condon and Lipton~\cite{CaiConLip94JCSS}
to prove that every language in $\PSPACE$
has a two-prover one-round interactive proof system
with an exponentially small one-sided error
in the classical world.
Therefore, an implication of \Ref\cite{ItoKobMat09CCC}
is that the protocol in \Ref\cite{CaiConLip94JCSS} has an unexpected strong soundness property:
the protocol remains to have an exponentially small error
even if we allow the two provers to behave arbitrarily
as long as they are no-signaling.

Given that the soundness analysis of protocols against no-signaling provers
is perhaps easier than that against entangled provers,
it is tempting to try to extend the result of \Ref\cite{ItoKobMat09CCC}
to a class of languages larger than $\PSPACE$.
For example, is it possible to construct
a two-prover one-round interactive proof system for $\NEXP$
which is sound against no-signaling provers?
The answer is no unless $\EXP=\NEXP$
because two-prover one-round interactive proof systems with no-signaling provers
can recognize at most $\EXP$
as pointed out by Preda~\cite{Preda}.
Then what about $\EXP$?

\subsection{Our results}

Let $\MIPns(2,1)$ be the class of languages
having a two-prover one-round interactive proof system
with no-signaling provers with bounded two-sided error.
The abovementioned result in \Ref\cite{ItoKobMat09CCC}
implies $\MIPns(2,1)\supseteq\PSPACE$.
Preda~\cite{Preda} shows $\MIPns(2,1)\subseteq\EXP$.

Our main result is:

\begin{theorem} \label{theorem:mipns-in-pspace}
  $\MIPns(2,1)\subseteq\PSPACE$.
\end{theorem}

An immediate corollary obtained by combining Theorem~\ref{theorem:mipns-in-pspace}
with the abovementioned result in \Ref\cite{ItoKobMat09CCC}
is the following exact characterization of the class~$\MIPns(2,1)$:

\begin{corollary}
  $\MIPns(2,1)=\PSPACE$,
  and this is achievable with exponentially small one-sided error,
  even if honest provers are restricted to be unentangled.
\end{corollary}

This puts the proof system of \Ref\cite{CaiConLip94JCSS}
in a rather special position:
while other two-prover one-round interactive proof systems~%
\cite{BabForLun91CC,Feige91SCT,FeiLov92STOC}
work with the whole $\NEXP$,
the one in \Ref\cite{CaiConLip94JCSS}
attains the best achievable
by two-prover one-round interactive proof systems with two-sided bounded error
that are sound against no-signaling provers,
and at the same time, it achieves an exponentially small one-sided error.

At a lower level, our result is actually a parallel algorithm
to approximately decide%
\footnote{The algorithm stated in Theorem~\ref{theorem:wns-in-nc}
  can be converted to an algorithm to approximate $\wns(G)$
  within an additive error in a standard way.
  See Remark~\ref{remark:decide-vs-compute} in Section~\ref{section:proof}.}
the value of a two-player one-round game for no-signaling players as follows.
For a two-player one-round game~$G$,
$\wns(G)$ is the value of $G$ for no-signaling provers
and $\abs{G}$ is the size of $G$,
both of which will be defined in Section~\ref{subsection:preliminaries-games}.

\begin{theorem} \label{theorem:wns-in-nc}
  There exists a parallel algorithm which,
  given a two-player one-round game~$G$
  and numbers $0\le s<c\le1$
  such that either $\wns(G)\le s$ or $\wns(G)\ge c$,
  decides which is the case.
  The parallel time of the algorithm is polynomial
  in $\log\abs{G}$ and $1/(c-s)$.
  The total work is polynomial
  in $\abs{G}$ and $1/(c-s)$.
\end{theorem}

Theorem~\ref{theorem:mipns-in-pspace} follows
by applying the algorithm of Theorem~\ref{theorem:wns-in-nc}
to the exponential-size game naturally arising
from a two-prover one-round interactive proof system.
This approach is similar to that of the recent striking result
on the $\PSPACE$ upper bound on $\QIP$~\cite{JaiJiUpaWat09}
as well as other complexity classes
related to quantum interactive proof systems,
i.e.\ $\QRG(1)$~\cite{JaiWat09CCC} and $\QIP(2)$~\cite{JaiUpaWat09FOCS}.%
\footnote{Do not be confused by an unfortunate inconsistency
  as for whether the number in the parenthesis represents
  the number of \emph{rounds} or \emph{turns},
  where one round consists of two turns.
  The ``$1$'' in $\QRG(1)$ and the ``$2$'' in $\QIP(2)$
  represent the number of turns
  whereas the ``$1$'' in $\MIPns(2,1)$ represents the number of rounds
  just in the same way as the ``$1$'' in $\MIP(2,1)$.}

The construction of the parallel algorithm in Theorem~\ref{theorem:wns-in-nc}
is much simpler than those used in \Refs\cite{JaiWat09CCC,JaiUpaWat09FOCS,JaiJiUpaWat09}
because our task can be formulated
as solving a \emph{linear} program of a certain special form approximately
instead of a \emph{semidefinite} program.
This allows us to use the fast parallel algorithm
for the mixed packing and covering problem by Young~\cite{Young01FOCS}.

\subsection{Organization of the paper}

The rest of this paper is organized as follows.
Section~\ref{section:preliminaries}
gives the definitions used later
and states the result by Young~\cite{Young01FOCS}
about a fast parallel approximation algorithm
for the mixed packing and covering problem.
Section~\ref{section:lemma} proves Theorem~\ref{theorem:mipns-in-pspace}
assuming Theorem~\ref{theorem:wns-in-nc}.
Section~\ref{section:proof} proves Theorem~\ref{theorem:wns-in-nc}
by using Young's fast parallel algorithm.
Section~\ref{section:conclusion} concludes the paper
by discussing some natural open problems.

\section{Preliminaries} \label{section:preliminaries}

We assume the familiarity with the notion
of multi-prover interactive proof systems.
Readers are referred to the textbook by Goldreich~\cite{Goldreich08}.

\subsection{Definitions on games} \label{subsection:preliminaries-games}

A protocol of a two-prover one-round interactive proof system
defines an exponential-size game for each instance.
Here we give a formal definition of \emph{games}.

A \emph{two-prover one-round game}, or simply a \emph{game} in this paper,
is played by two cooperative provers called the prover~1 and the prover~2
with help of a verifier who enforces the rule of the game.
A game is formulated as $G=(Q_1,Q_2,A_1,A_2,\pi,R)$
by nonempty finite sets~$Q_1$, $Q_2$, $A_1$ and $A_2$,
a probability distribution~$\pi$ over $Q_1\times Q_2$,
and a function $R\colon Q_1\times Q_2\times A_1\times A_2\to[0,1]$.
As is customary, we write $R(q_1,q_2,a_1,a_2)$ as $R(a_1,a_2\mid q_1,q_2)$.

In this game,
the verifier generates a pair of questions $(q_1,q_2)\in Q_1\times Q_2$
according to the probability distribution $\pi$,
and sends $q_1$ to the prover~1 and $q_2$ to the prover~2.
Each prover~$\nu$ ($\nu\in\{1,2\}$) sends an answer~$a_\nu\in A_\nu$ to the verifier
without knowing the question sent to the other prover.
Finally, the verifier accepts with probability~$R(a_1,a_2\mid q_1,q_2)$
and rejects with probability~$1-R(a_1,a_2\mid q_1,q_2)$.
The provers try to make the verifier accept with as high probability as possible.

The \emph{size}~$\abs{G}$ of the game~$G$ is defined as
$\abs{G}=\abs{Q_1}\abs{Q_2}\abs{A_1}\abs{A_2}$.

A \emph{strategy} in a two-prover one-round game~$G$
is a family~$p=(p_{q_1q_2})$ of probability distributions on~$A_1\times A_2$
indexed by $(q_1,q_2)\in Q_1\times Q_2$.
As is customary, the probability~$p_{q_1q_2}(a_1,a_2)$
is written as $p(a_1,a_2\mid q_1,q_2)$.
A strategy~$p$ is said to be \emph{no-signaling}
if it satisfies the following \emph{no-signaling conditions}:
\begin{itemize}
\item
  The marginal probability~$p_1(a_1\mid q_1)=\sum_{a_2}p(a_1,a_2\mid q_1,q_2)$ does not depend on~$q_2$.
\item
  The marginal probability~$p_2(a_2\mid q_2)=\sum_{a_1}p(a_1,a_2\mid q_1,q_2)$ does not depend on~$q_1$.
\end{itemize}

The \emph{acceptance probability} of a strategy~$p$ is given by
\begin{align*}
  \sum_{q_1\in Q_1,q_2\in Q_2}\pi(q_1,q_2)
  \sum_{a_1\in A_1,a_2\in A_2}R(a_1,a_2\mid q_1,q_2)p(a_1,a_2\mid q_1,q_2).
\end{align*}
The \emph{no-signaling value} $\wns(G)$ of $G$
is the maximum of the acceptance probability over all no-signaling strategies.

\subsection{Definitions on interactive proof systems}

Let $\Sigma=\{0,1\}$.
A \emph{two-prover one-round interactive proof system}
is defined by a polynomial~$l\colon\ZZ_{\ge0}\to\ZZ_{\ge0}$,
a polynomial-time computable mapping~%
$M_\pi\colon\Sigma^*\times\Sigma^*\to\Sigma^*\times\Sigma^*$
such that $x\in\Sigma^n$ and $r\in\Sigma^{l(n)}$
imply $M_\pi(x,r)\in\Sigma^{l(n)}\times\Sigma^{l(n)}$,
and a polynomial-time decidable predicate~%
$M_R\colon\Sigma^*\times\Sigma^*\times\Sigma^*\times\Sigma^*\to\{0,1\}$.
On receiving an input string~$x\in\Sigma^*$,
the verifier prepares an $l(\abs{x})$-bit string~$r$ uniformly at random
and computes $(q_1,q_2)=M_\pi(x,r)$.
Then he sends each string~$q_\nu$ ($\nu\in\{1,2\}$) to the prover~$\nu$
and receives an $l(\abs{x})$-bit string~$a_\nu$ from each prover~$\nu$.
Finally he accepts if and only if $M_R(x,r,a_1,a_2)=1$.
This naturally defines a game~%
$G^{(x)}=(Q_1^{(x)},Q_2^{(x)},A_1^{(x)},A_2^{(x)},\pi^{(x)},R^{(x)})$
for each input string~$x$,
where $Q_1^{(x)}=Q_2^{(x)}=A_1^{(x)}=A_2^{(x)}=\Sigma^{l(\abs{x})}$,
\begin{align*}
  \pi^{(x)}(q_1,q_2)&=2^{-l(\abs{x})}\cdot
    \#\{r\in\Sigma^{l(\abs{x})}\mid M_\pi(x,r)=(q_1,q_2)\}, \\
  R^{(x)}(a_1,a_2\mid q_1,q_2)&=
    \frac{\#\{r\in\Sigma^{l(\abs{x})}\mid M_\pi(x,r)=(q_1,q_2)\wedge M_R(x,r,a_1,a_2)=1\}}
         {\#\{r\in\Sigma^{l(\abs{x})}\mid M_\pi(x,r)=(q_1,q_2)\}}.
\end{align*}

Let $c,s\colon\ZZ_{\ge0}\to[0,1]$ be functions such that $c(n)>s(n)$ for every $n$.
The two-prover one-round interactive proof system
is said to \emph{recognize a language%
\footnote{Although we define $\MIPns(2,1)$ as a class of languages in this paper
  to keep the notations simple,
  we could alternatively define $\MIPns(2,1)$ as the class of
  \emph{promise problems}~\cite{EveSelYac84IC,Goldreich05-promise}
  recognized by a two-prover one-round interactive proof system
  with no-signaling provers.
  A generalization of Theorem~\ref{theorem:mipns-in-pspace}
  to the case of promise problems is straightforward.}%
~$L$ with completeness acceptance probability at least~$c(n)$
and soundness error at most~$s(n)$
with no-signaling provers}
when the following conditions are satisfied.
\begin{description}
\item[Completeness]
  $x\in L\implies\wns(G^{(x)})\ge c(\abs{x})$.
\item[Soundness]
  $x\notin L\implies\wns(G^{(x)})\le s(\abs{x})$.
\end{description}

In particular, the proof system is said to
\emph{recognize $L$ with bounded errors with no-signaling provers}
if the binary representations of $c(n)$ and $s(n)$
are computable in time polynomial in $n$
and there exists a polynomial~$f\colon\ZZ_{\ge0}\to\ZZ_{\ge1}$
such that for every $n$, it holds $c(n)-s(n)>1/f(n)$.
We denote by $\MIPns(2,1)$
the class of languages~$L$
which are recognized by a two-prover one-round interactive proof system
with bounded errors
with no-signaling provers.

\subsection{Mixed packing and covering problem}

The \emph{mixed packing and covering problem}
is the linear feasibility problem of the form
\begin{alignat*}{2}
  &\text{Find} &\quad
    &x\in\RR^N, \\
  &\text{Such that} &\quad
    &Ax\le b, \\
  &&&Cx\ge d, \\
  &&&x\ge0,
\end{alignat*}
where matrices~$A,C$ and vectors~$b,d$ are given
and the entries of $A,b,C,d$ are all nonnegative.
For $r\ge1$,
an \emph{$r$-approximate solution} is a vector~$x\ge0$
such that $Ax\le rb$ and $Cx\ge d$.

\begin{theorem}[Young~\cite{Young01FOCS}]
  \label{theorem:Young01FOCS}
  There exists a parallel algorithm
  which, given an instance~$(A,b,C,d)$ of the mixed packing and covering problem
  and a number~$\varepsilon>0$,
  either:
  \begin{itemize}
  \item
    claims that the given instance does not have a feasible solution, or
  \item
    finds a $(1+\varepsilon)$-approximate solution.
  \end{itemize}
  If the size of $A$ and $C$ are $M_1\times N$ and $M_2\times N$, respectively,
  then the algorithm runs in parallel time polynomial
  in~$\log M_1$, $\log M_2$, $\log N$ and $1/\varepsilon$
  and total work polynomial
  in~$M_1$, $M_2$, $N$ and $1/\varepsilon$.
\end{theorem}

\section{Proof of Theorem~\ref{theorem:mipns-in-pspace}} \label{section:lemma}

Theorem~\ref{theorem:mipns-in-pspace} follows from Theorem~\ref{theorem:wns-in-nc}
by a standard argument
using the polynomial equivalence
between space and parallel time~\cite{Borodin77SICOMP}.

Let $L\in\MIPns(2,1)$,
and fix an two-prover one-round interactive proof system
which recognizes $L$ with bounded errors with no-signaling provers.
Let $c(n)$ and $s(n)$ be the completeness acceptance probability
and the soundness error of this proof system, respectively.
We construct a polynomial-space algorithm which recognizes $L$.

Let $x$ be an input string and $n=\abs{x}$.
Let
$G^{(x)}=(Q_1^{(x)},\allowbreak Q_2^{(x)},\allowbreak
          A_1^{(x)},\allowbreak A_2^{(x)},\allowbreak
          \pi^{(x)},\allowbreak R^{(x)})$
be the game naturally arising from the proof system on input~$x$.
The size of $Q_1^{(x)},Q_2^{(x)},A_1^{(x)},A_2^{(x)}$
is at most exponential in $n$.
For each $q_1,q_2,a_1,a_2$, it is possible to compute
$\pi^{(x)}(q_1,q_2)$ and $R^{(x)}(a_1,a_2\mid q_1,q_2)$
in space polynomial in $n$
by simulating every choice of randomness of the verifier.
By Theorem~4 of Borodin~\cite{Borodin77SICOMP},
the parallel algorithm of Theorem~\ref{theorem:wns-in-nc}
can be converted to a sequential algorithm
which runs in space polynomial
in $\log\abs{G}$ and $1/(c-s)$.
By applying this algorithm to the game~$G^{(x)}$
and the threshold values~$c(\abs{x})$ and $s(\abs{x})$,
we decide whether $\wns(G^{(x)})\ge c(\abs{x})$ or $\wns(G^{(x)})\le s(\abs{x})$,
or equivalently whether $x\in L$ or $x\notin L$,
in space polynomial in~$\log\abs{G^{(x)}}=\poly(n)$
and~$1/(c(\abs{x})-s(\abs{x}))=\poly(n)$.

Note that the composition of two functions computable in space polynomial in $\abs{x}$
is also computable in space polynomial in $\abs{x}$,
which can be proved in the same way as Proposition~8.2 of \Ref\cite{Papadimitriou94}.

\section{Formulating no-signaling value by mixed packing and covering problem} \label{section:proof}

This section proves Theorem~\ref{theorem:wns-in-nc}.

Let $G=(Q_1,Q_2,A_1,A_2,\pi,R)$ be a game.
Let $\pi_1(q_1)=\sum_{q_2\in Q_2}\pi(q_1,q_2)$
and $\pi_2(q_2)=\sum_{q_1\in Q_1}\pi(q_1,q_2)$ be the marginal distributions.
Without loss of generality,
we assume that every question in $Q_1$ and $Q_2$ is used with nonzero probability,
i.e.\ $\pi_1(q_1)>0$ for every $q_1\in Q_1$ and $\pi_2(q_2)>0$ for every $q_2\in Q_2$.

By definition, the no-signaling value $\wns(G)$ of $G$ is equal to
the optimal value of the following linear program:
\begin{subequations}
  \label{eq:no-sig}
\begin{alignat}{3}
  &\text{Maximize} &\quad
  &\sum_{q_1,q_2}\pi(q_1,q_2)\sum_{a_1,a_2}R(a_1,a_2\mid q_1,q_2)p(a_1,a_2\mid q_1,q_2),
    \hspace*{-5em} &&
    \label{eq:no-sig-obj} \\
  &\text{Subject to} &\quad
  &\sum_{a_2}p(a_1,a_2\mid q_1,q_2)=p_1(a_1\mid q_1), &\quad
    &\forall q_1\in Q_1,\;q_2\in Q_2,\;a_1\in A_1,
    \label{eq:no-sig-con-1} \\
  &&&\sum_{a_1}p(a_1,a_2\mid q_1,q_2)=p_2(a_2\mid q_2), &\quad
    &\forall q_1\in Q_2,\;q_2\in Q_2,\;a_2\in A_2,
    \label{eq:no-sig-con-2} \\
  &&&\sum_{a_1,a_2}p(a_1,a_2\mid q_1,q_2)=1, &\quad
    &\forall q_1\in Q_1,\;q_2\in Q_2,
    \label{eq:no-sig-con-3} \\
  &&&p(a_1,a_2\mid q_1,q_2)\ge0, &\quad
    &\forall q_1\in Q_1,\;q_2\in Q_2,\;a_1\in A_1,\;a_2\in A_2.
    \label{eq:no-sig-con-4}
\end{alignat}
\end{subequations}

We transform this linear program~(\ref{eq:no-sig}) successively
without changing the optimal value.
First, we replace the constraint~(\ref{eq:no-sig-con-3}) by two constraints:
\begin{subequations}
\begin{alignat}{2}
  &\sum_{a_1}p_1(a_1\mid q_1)=1, &\quad
    &\forall q_1\in Q_1,
    \label{eq:no-sig-con-3-1} \\
  &\sum_{a_2}p_2(a_2\mid q_2)=1, &\quad
    &\forall q_2\in Q_2.
    \label{eq:no-sig-con-3-2}
\end{alignat}
\end{subequations}
It is clear that this rewriting does not change the optimal value.

Next, we relax the constraints~(\ref{eq:no-sig-con-1})
and (\ref{eq:no-sig-con-2}) to inequalities:
\begin{subequations}
\begin{alignat}{2}
  &\sum_{a_2}p(a_1,a_2\mid q_1,q_2)\le p_1(a_1\mid q_1), &\quad
    &\forall q_1\in Q_1,\;q_2\in Q_2,\;a_1\in A_1,
    \label{eq:no-sig-con-1-ineq} \\
  &\sum_{a_1}p(a_1,a_2\mid q_1,q_2)\le p_2(a_2\mid q_2), &\quad
    &\forall q_1\in Q_1,\;q_2\in Q_2,\;a_2\in A_2.
    \label{eq:no-sig-con-2-ineq}
\end{alignat}
\end{subequations}

\begin{claim} \label{claim:inequalities}
  The optimal value~$w$ of the linear program~(\ref{eq:no-sig})
  is equal to the maximum value~$w'$ of (\ref{eq:no-sig-obj})
  subject to the constraints~(\ref{eq:no-sig-con-4}),
  (\ref{eq:no-sig-con-3-1}), (\ref{eq:no-sig-con-3-2}),
  (\ref{eq:no-sig-con-1-ineq}) and (\ref{eq:no-sig-con-2-ineq}).
\end{claim}

\begin{proof}
  Since we only relaxed the constraints, $w\le w'$ is obvious.
  To prove $w\ge w'$, let $(\tilde{p},p_1,p_2)$
  be a solution satisfying
  the constraints~(\ref{eq:no-sig-con-4}),
  (\ref{eq:no-sig-con-3-1}), (\ref{eq:no-sig-con-3-2}),
  (\ref{eq:no-sig-con-1-ineq}) and (\ref{eq:no-sig-con-2-ineq}).
  We will construct $p$
  such that $(p,p_1,p_2)$ is a feasible solution of the linear program~(\ref{eq:no-sig})
  and $p(a_1,a_2\mid q_1,q_2)\ge\tilde{p}(a_1,a_2\mid q_1,q_2)$ for every $q_1,q_2,a_1,a_2$.

  Fix any $q_1,q_2\in Q$.
  Let
  \begin{alignat*}{2}
    s_{q_1q_2}(a_1)&=p_1(a_1\mid q_1)-\sum_{a_2\in A_2}\tilde{p}(a_1,a_2\mid q_1,q_2), &\quad
      &\forall a_1\in A_1, \\
    t_{q_1q_2}(a_2)&=p_2(a_2\mid q_2)-\sum_{a_1\in A_1}\tilde{p}(a_1,a_2\mid q_1,q_2), &\quad
      &\forall a_2\in A_2.
  \end{alignat*}
  The following relations are easy to verify:
  \begin{subequations}
  \begin{align}
    &s_{q_1q_2}(a_1)\ge0, \quad \forall a_1\in A_1,
      \label{eq:dominated-cond-3} \\
    &t_{q_1q_2}(a_2)\ge0, \quad \forall a_2\in A_2,
      \label{eq:dominated-cond-4} \\
    &\sum_{a_1\in A_1}s_{q_1q_2}(a_1)=\sum_{a_2\in A_2}t_{q_1q_2}(a_2) \;
      (=:F_{q_1q_2}).
      \label{eq:dominated-cond-5}
  \end{align}
  \end{subequations}
  We define $p(a_1,a_2\mid q_1,q_2)$ by
  \[
    p(a_1,a_2\mid q_1,q_2)=\begin{cases}
      \tilde{p}(a_1,a_2\mid q_1,q_2)+\frac{1}{F_{q_1q_2}}s_{q_1q_2}(a_1)t_{q_1q_2}(a_2), &
        \text{if } F_{q_1q_2}>0, \\
      \tilde{p}(a_1,a_2\mid q_1,q_2), &
        \text{if } F_{q_1q_2}=0.
    \end{cases}
  \]
  Then it is clear
  from Eqs.~(\ref{eq:dominated-cond-3}) and (\ref{eq:dominated-cond-4})
  that $p(a_1,a_2\mid q_1,q_2)\ge\tilde{p}(a_1,a_2\mid q_1,q_2)$
  for every $q_1,q_2,a_1,a_2$.
  Eqs.~(\ref{eq:no-sig-con-1}) and (\ref{eq:no-sig-con-2})
  follow from Eq.~(\ref{eq:dominated-cond-5}).
\end{proof}

Replace the variables $p(a_1,a_2\mid q_1,q_2)$
by $x(a_1,a_2\mid q_1,q_2)=\pi(q_1,q_2)p(a_1,a_2\mid q_1,q_2)$.
The resulting linear program is as follows.
\begin{subequations}
  \label{eq:lp-1}
\begin{alignat}{3}
  &\text{Maximize} &\quad
    &\sum_{a_1,a_2,q_1,q_2}R(a_1,a_2\mid q_1,q_2)x(a_1,a_2\mid q_1,q_2), && \label{eq:lp-1-obj} \\
  &\text{Subject to} &\quad
    &\sum_{a_2} x(a_1,a_2\mid q_1,q_2)\le \pi(q_1,q_2)p_1(a_1\mid q_1), &\quad
    &\forall q_1\in Q_1,\;q_2\in Q_2,\;a_1\in A_1, \label{eq:lp-1-con-1} \\
  &&&\sum_{a_1} x(a_1,a_2\mid q_1,q_2)\le \pi(q_1,q_2)p_2(a_2\mid q_2), &\quad
    &\forall q_1\in Q_1,\;q_2\in Q_2,\;a_2\in A_2, \label{eq:lp-1-con-2} \\
  &&&\sum_{a_1} p_1(a_1\mid q_1)=1, &\quad
    &\forall q_1\in Q_1, \label{eq:lp-1-con-3} \\
  &&&\sum_{a_2} p_2(a_2\mid q_2)=1, &\quad
    &\forall q_2\in Q_2, \label{eq:lp-1-con-4} \\
  &&&x(a_1,a_2\mid q_1,q_2)\ge0, &\quad
    &\forall q_1\in Q_1,\;q_2\in Q_2,\;a_1\in A_1,\;a_2\in A_2. \label{eq:lp-1-con-5}
\end{alignat}
\end{subequations}

By the strong duality theorem of linear programming,
the linear program~(\ref{eq:lp-1}) has the same objective value
as the following:
\begin{subequations}
  \label{eq:lp-1-dual}
\begin{alignat}{3}
  &\text{Minimize} &\quad
  &\sum_{q_1}z_1(q_1)+\sum_{q_2}z_2(q_2), && \label{eq:lp-1-dual-obj} \\
  &\text{Subject to} &\quad
  &y_1(q_1,q_2,a_1)+y_2(q_1,q_2,a_2)\ge R(a_1,a_2\mid q_1,q_2), \hspace*{-10em} && \nonumber \\
  &&&&\quad &\forall q_1\in Q_1,\;q_2\in Q_2,\;a_1\in A_1,\;a_2\in A_2, \label{eq:lp-1-dual-con-1} \\
  &&&z_1(q_1)\ge\sum_{q_2}\pi(q_1,q_2)y_1(q_1,q_2,a_1), &\quad
    &\forall q_1\in Q_1,\;a_1\in A_1, \label{eq:lp-1-dual-con-2} \\
  &&&z_2(q_2)\ge\sum_{q_1}\pi(q_1,q_2)y_2(q_1,q_2,a_2), &\quad
    &\forall q_2\in Q_2,\;a_2\in A_2, \label{eq:lp-1-dual-con-3} \\
  &&&y_1(q_1,q_2,a_1)\ge0, &\quad
    &\forall q_1\in Q_1,\;q_2\in Q_2,\;a_1\in A_1, \label{eq:lp-1-dual-con-4} \\
  &&&y_2(q_1,q_2,a_2)\ge0, &\quad
    &\forall q_1\in Q_1,\;q_2\in Q_2,\;a_2\in A_2. \label{eq:lp-1-dual-con-5}
\end{alignat}
\end{subequations}

Note that the constraints~(\ref{eq:lp-1-dual-con-2})--(\ref{eq:lp-1-dual-con-5})
imply $z_1(q_1)\ge0$ and $z_2(q_2)\ge0$.

Let $(z_1,z_2,y_1,y_2)$ be a feasible solution
of the linear program~(\ref{eq:lp-1-dual}).
If $y_1(q_1,q_2,a_1)>1$ for some $q_1,q_2,a_1$,
we can replace $y_1(q_1,q_2,a_1)$ by $1$
without violating any constraints or increasing the objective value.
The same holds for $y_2(q_1,q_2,a_2)$.
Therefore, adding the constraints
\begin{alignat*}{2}
  &y_1(q_1,q_2,a_1)\le1, &\quad
    &\forall q_1\in Q_1,\;q_2\in Q_2,\;a_1\in A_1, \\
  &y_2(q_1,q_2,a_2)\le1, &\quad
    &\forall q_1\in Q_1,\;q_2\in Q_2,\;a_2\in A_2
\end{alignat*}
does not change the optimal value.

Replacing the variables~$y_1(q_1,q_2,a_1)$ by $1-\bar{y}_1(q_1,q_2,a_1)$
and $y_2(q_1,q_2,a_2)$ by $1-\bar{y}_2(q_1,q_2,a_2)$,
the following claim is immediate.

\begin{claim} \label{claim:positive-coeffs}
  The no-signaling value~$\wns(G)$ is equal to the optimal value of the following linear program.
  \begin{subequations}
    \label{eq:lp-2}
  \begin{alignat}{3}
    &\text{Minimize} &\quad
    &\sum_{q_1}z_1(q_1)+\sum_{q_2}z_2(q_2), && \label{eq:lp-2-obj} \\
    &\text{Subject to} &\quad
    &\bar{y}_1(q_1,q_2,a_1)+\bar{y}_2(q_1,q_2,a_2)\le 2-R(a_1,a_2\mid q_1,q_2), \hspace*{-10em} && \nonumber \\
    &&&&\quad &\forall q_1\in Q_1,\;q_2\in Q_2,\;a_1\in A_1,\;a_2\in A_2, \label{eq:lp-2-con-1} \\
    &&&z_1(q_1)+\sum_{q_2}\pi(q_1,q_2)\bar{y}_1(q_1,q_2,a_1)\ge\pi_1(q_1), &\quad
      &\forall q_1\in Q_1,\;a_1\in A_1, \label{eq:lp-2-con-2} \\
    &&&z_2(q_2)+\sum_{q_1}\pi(q_1,q_2)\bar{y}_2(q_1,q_2,a_2)\ge\pi_2(q_2), &\quad
      &\forall q_2\in Q_2,\;a_2\in A_2, \label{eq:lp-2-con-3} \\
    &&&\bar{y}_1(q_1,q_2,a_1)\le1, &\quad
      &\forall q_1\in Q_1,\;q_2\in Q_2,\;a_1\in A_1, \label{eq:lp-2-con-4} \\
    &&&\bar{y}_2(q_1,q_2,a_2)\le1, &\quad
      &\forall q_1\in Q_1,\;q_2\in Q_2,\;a_2\in A_2, \label{eq:lp-2-con-5} \\
    &&&\bar{y}_1(q_1,q_2,a_1)\ge0, &\quad
      &\forall q_1\in Q_1,\;q_2\in Q_2,\;a_1\in A_1, \label{eq:lp-2-con-6} \\
    &&&\bar{y}_2(q_1,q_2,a_2)\ge0, &\quad
      &\forall q_1\in Q_1,\;q_2\in Q_2,\;a_2\in A_2, \label{eq:lp-2-con-7} \\
    &&&z_1(q_1)\ge0, &\quad
      &\forall q_1\in Q_1, \label{eq:lp-2-con-8} \\
    &&&z_2(q_2)\ge0, &\quad
      &\forall q_2\in Q_2. \label{eq:lp-2-con-9}
  \end{alignat}
  \end{subequations}
\end{claim}

\begin{lemma} \label{lemma:packing-covering}
  Let $G=(Q_1,Q_2,A_1,A_2,\pi,R)$ be a game and $0\le s<c\le1$.
  Consider the instance of the mixed packing and covering problem
  consisting of a constraint~$\sum_{q_1}z_1(q_1)+\sum_{q_2}z_2(q_2)\le s$
  and the constraints~(\ref{eq:lp-2-con-1})--(\ref{eq:lp-2-con-9}).
  Let $\varepsilon=(c-s)/4$.
  Then,
  \begin{enumerate}[(i)]
  \item
    If $\wns(G)\le s$, this instance has a feasible solution.
  \item
    If $\wns(G)\ge c$, this instance does not have a $(1+\varepsilon)$-approximate solution.
  \end{enumerate}
\end{lemma}

\begin{proof}
  \begin{enumerate}[(i)]
  \item
    Clear from Claim~\ref{claim:positive-coeffs}.
  \item
    We prove the contrapositive.
    Assume that $(\bar{y}_1,\bar{y}_2,z_1,z_2)$ is a $(1+\varepsilon)$-approximate solution,
    and let
    \begin{align*}
      \bar{y}'_1(q_1,q_2,a_1)&=\frac{1}{1+\varepsilon}\bar{y}_1(q_1,q_2,a_1), \\
      \bar{y}'_2(q_1,q_2,a_2)&=\frac{1}{1+\varepsilon}\bar{y}_2(q_1,q_2,a_2), \\
      z'_1(q_1)&=z_1(q_1)+\varepsilon\pi_1(q_1), \\
      z'_2(q_2)&=z_2(q_2)+\varepsilon\pi_2(q_2).
    \end{align*}
    Then $(\bar{y}'_1,\bar{y}'_2,z'_1,z'_2)$
    satisfies (\ref{eq:lp-2-con-1})--(\ref{eq:lp-2-con-9}),
    and
    \[
      \sum_{q_1}z'_1(q_1)+\sum_{q_2}z'_2(q_2)
      =\sum_{q_1}z_1(q_1)+\varepsilon\sum_{q_1}\pi_1(q_1)+\sum_{q_2}z_2(q_2)+\varepsilon\sum_{q_2}\pi_2(q_2)
      \le s+3\varepsilon<c.
    \]
    Therefore, $\wns(G)$,
    or the optimal value of the linear program~(\ref{eq:lp-2}),
    is less than $c$.
    \qedhere
  \end{enumerate}
\end{proof}

\begin{proof}[Proof of Theorem~\ref{theorem:wns-in-nc}]
  Apply Theorem~\ref{theorem:Young01FOCS}
  to the instance of the mixed packing and covering problem
  in Lemma~\ref{lemma:packing-covering}.
\end{proof}

\begin{remark}
  It is easy to see that adding the constraints~$z_1(q_1)\le\pi_1(q_1)$ for~$q_1\in Q_1$
  and $z_2(q_2)\le\pi_2(q_2)$ for~$q_2\in Q_2$
  to the instance of the mixed packing and covering problem
  in Lemma~\ref{lemma:packing-covering}
  does not change the feasibility or approximate feasibility.
  The resulting linear program has a constant ``width''
  in the sense stated in Theorem~2.12
  of Plotkin, Shmoys and Tardos~\cite{PloShmTar95MOR}
  with a suitable tolerance vector.
  See \cite{PloShmTar95MOR} for relevant definitions.
  This gives an alternative proof of Theorem~\ref{theorem:wns-in-nc}
  which uses the algorithm of \cite{PloShmTar95MOR}
  instead of the algorithm of \cite{Young01FOCS}.
\end{remark}

\begin{remark} \label{remark:decide-vs-compute}
  Given Theorem~\ref{theorem:wns-in-nc},
  it is easy to approximate $\wns(G)$ within additive error~$\varepsilon$
  (rather than deciding whether $\wns(G)\le s$ or $\wns(G)\ge c$)
  in parallel time polynomial in $\log\abs{G}$ and $1/\varepsilon$
  and total work polynomial in $\abs{G}$ and $1/\varepsilon$.
  This can be done by trying all the possibilities of
  $s=k\varepsilon$ and $c=(k+1)\varepsilon$
  for integers~$k$ in the range~$0\le k\le 1/\varepsilon$
  in parallel,
  or by using the binary search.
\end{remark}

\section{Concluding remarks} \label{section:conclusion}

This paper gave the exact characterization
of the simplest case of multi-prover interactive proof systems
with no-signaling provers: $\MIPns(2,1)=\PSPACE$.
A natural direction seems to be to extend this result
to show a $\PSPACE$ upper bound on a class containing $\MIPns(2,1)$.
Below we discuss some hurdles in doing so.
\begin{itemize}
\item
  \emph{More than two provers.}
  In the completely classical case,
  a many-prover one-round interactive proof system
  can be transformed to a two-prover one-round interactive proof system
  by using the oracularization technique,
  and therefore $\MIP(\poly,1)\subseteq\MIP(2,1)$.
  The same transformation is not known to preserve soundness
  in the case of no-signaling provers
  even when the original proof system uses three provers.%
  \footnote{The Magic Square game in \Ref\cite{CleHoyTonWat04CCC}
    is a counterexample which shows that
    this transformation cannot be used alone
    to reduce the number of provers from three to two
    in the case of \emph{entangled} provers
    because it sometimes transforms a three-prover game
    whose entangled value is less than $1$
    to a two-prover game
    whose entangled value is equal to $1$~\cite{ItoKobPreSunYao08CCC}.
    The situation might be different
    in the case of no-signaling provers.}
  As a result, whether or not $\MIPns(3,1)\subseteq\MIPns(2,1)$ is unknown,
  and our result does not imply $\MIPns(3,1)\subseteq\PSPACE$.

  To extend the current proof to $\MIPns(3,1)$,
  the main obstacle is to extend Claim~\ref{claim:inequalities},
  which replaces equations by inequalities.
  It does not seem that an analogous claim can be proved for three provers
  by a straightforward extension of the current proof of Claim~\ref{claim:inequalities}.
\item
  \emph{More than one round.}
  The proof of Claim~\ref{claim:inequalities} seems to work
  in the case of two-prover systems with polynomially many rounds.
  However, in a linear program corresponding to (\ref{eq:lp-1-dual}),
  an upper bound on the variables $y_1$ and $y_2$
  becomes exponentially large
  and the current proof does not work
  even in the case of two-prover two-round systems with adaptive questions
  or two-prover $\omega(\log n)$-round systems with non-adaptive questions.
\item
  \emph{Quantum verifier and quantum messages.}
  The notion of no-signaling strategies
  can be extended to the case of quantum messages~\cite{BecGotNiePre01PRA,Gutoski09QIC}
  (\Ref\cite{BecGotNiePre01PRA} uses the term ``causal'' instead of ``no-signaling'').
  This allows us to define e.g.\ the class~$\QMIPns(2,2)$ of languages
  having a \emph{quantum} two-prover one-round (two-turn)
  interactive proof system with no-signaling provers.
  The class~$\QMIPns(2,2)$ contains both $\MIPns(2,1)$ and $\QIP(2)$,
  and it would be nice if the method of \Ref\cite{JaiUpaWat09FOCS} and ours
  can be unified to give $\QMIPns(2,2)=\PSPACE$.
  One obvious obstacle is how to extend the fast parallel algorithm
  in \Ref\cite{JaiUpaWat09FOCS}
  for the special case of semidefinite programming
  to the case of $\QMIPns(2,2)$.
  Another obstacle is again Claim~\ref{claim:inequalities};
  the current proof of Claim~\ref{claim:inequalities}
  essentially constructs of a joint probability distribution
  over $(q_1,q_2,a_1,a_2)$
  from its marginal distributions over $(q_1,q_2,a_1)$ and $(q_1,q_2,a_2)$,
  and this kind of \emph{state extension} is not always possible
  in the quantum case~\cite{Werner89-StateExtension,Werner90-StateExtension}.
\end{itemize}

\section*{Acknowledgment}

The author thanks Rahul Jain, Julia Kempe, Hirotada Kobayashi, Sarvagya Upadhyay
and John Watrous for helpful discussions.

\end{document}